\documentclass{article}%
\usepackage{graphicx}
\usepackage{amsmath}%
\usepackage{amsfonts}%
\usepackage{amssymb}
\newtheorem{theorem}{Theorem}
\newtheorem{acknowledgement}[theorem]{Acknowledgement}

\begin{document}

\title{Resonances, Unstable Systems and Irreversibility: Matter Meets Mind}
\author{Robert C. Bishop$^{a,b}$\\$^{a}$Center for Junior Research Fellows\\Box M682, Universit\"{a}t Konstanz, D-78457\\$^{b}$Faculty of Philosophy, University of Oxford\\Oxford OX1 4JJ, United Kingdom}
\date{}
\maketitle

\begin{abstract}
The fundamental time-reversal invariance of dynamical systems can be broken in
various ways. One way is based on the presence of resonances and their
interactions giving rise to unstable dynamical systems, leading to
well-defined time arrows. Associated with these time arrows are semigroups
bearing time orientations. Usually, when time symmetry is broken, two
time-oriented semigroups result, one directed toward the future and one
directed toward the past. If time-reversed states and evolutions are excluded
due to resonances, then the status of these states and their associated
backwards-in-time oriented semigroups is open to question. One possible role
for these latter states and semigroups is as an abstract representation of
mental systems as opposed to material systems. The beginnings of this
interpretation will be sketched.

\end{abstract}

\section{\noindent Introduction}

Usually dynamical systems are considered to be time-reversible as their
equations of motion are time-reversal symmetric under the time inversion
operator $R:(\vec{x},t)\rightarrow(\vec{x},-t)$. This means that if $\phi(t)$
is a solution of the equations of motion, then so is $R\phi(t)$. Such systems
should then be reversible in the sense that if they exhibit a temporal
succession of state transitions $\phi_{1}$, $\phi_{2}$, $\phi_{3}$,...,
$\phi_{n}$, they can also exhibit the reverse temporal sequence $R\phi_{n}$,
$R\phi_{n-1}$, $R\phi_{n-2}$,..., $R\phi_{1}$. In quantum mechanics these
evolutions typically are described by one-parameter unitary groups of operators.

Resonances appear in a number of dynamical systems, both classical and quantum
(e.g., Antoniou and Prigogine 1993; Antoniou and Tasaki 1993; Bohm et al.
1997) and are prototypical irreversible processes (e.g. scattering
resonances). When the number of resonances in dynamical systems is
sufficiently large, the dynamics is extremely unstable (e.g., exhibiting
sensitive dependence on initial conditions), and becomes irreversible. For
such unstable systems time arrows for the dynamics can be clearly defined
(e.g., Bishop 2004a; Bishop 2004b; Bishop 2005a). In the rigged Hilbert space
framework for quantum mechanics (e.g., Bohm and Gadella 1989; Bohm et al.
1997), such time arrows are represented by semigroups. It is typically the
case that there are two possible semigroups for such dynamics, one defined in
the forward direction in time and one defined in the backward direction.
However, if the evolutions of such resonance phenomena as scattering
resonances and quasistable particles are irreversible, then there appears to
be no physical relevance to the mathematical descriptions of the time-reversed
states and evolutions.

After presenting the background of the rigged Hilbert space (RHS) framework
for quantum mechanics (QM) in section 2, I will review its application to
resonance states for scattering (section 3). This will be followed by a brief
review of the extended Galilean group of Wigner and its application to
resonance states in the RHS framework (section 4). I will then give an
interpretation of the time-reversed resonance states and evolutions as
abstract representations of mental systems as opposed to material systems
(section 5).

\section{Rigged Hilbert Space Quantum Mechanics}

An RHS may be briefly characterized as follows. Let $\Psi$ be an abstract
linear scalar product space and complete $\Psi$ with respect to two
topologies. The first topology is the standard Hilbert space (HS) topology
$\tau_{\mathcal{H}}$ defined by the norm%
\begin{equation}
\Vert h\Vert=\sqrt{(h,h)}%
\end{equation}
where $h$ is an element of $\Psi$. The second topology $\tau_{\Phi}$ is
defined by a countable set of norms%
\begin{equation}
\Vert\phi\Vert_{n}=\sqrt{(\phi,\phi)_{n}},\;n=0,1,2,...
\end{equation}
where $\phi$ is also an element of $\Psi$ and the scalar product in (2) is
given by
\begin{equation}
(\phi,\phi^{\prime})_{n}=(\phi,(\Delta+1)^{n}\phi^{\prime}),\;n=0,1,2,...
\end{equation}
where $\Delta$ is the Nelson operator $\Delta=$ $\sum_{i}\chi_{i}^{2}$. The
$\chi_{i}$ are the generators of an enveloping algebra of observables for the
system in question and they form a basis for a Lie algebra (Nelson 1959; Bohm
et al. 1999). In the case of the harmonic oscillator, for example, the
$\chi_{i}$ would be the position and momentum operators or, alternatively, the
raising and lowering operators. Furthermore if the operator $\Delta+1$ is
\textit{nuclear }then the space $\Phi$ defined by (2) is a nuclear space
(Treves 1967).

A Gel'fand triplet is obtained by completing $\Psi$ with respect to
$\tau_{\Phi}$ to obtain $\Phi$ and with respect to $\tau_{\mathcal{H}}$ to
obtain $\mathcal{H}$. In addition there are the dual spaces of continuous
linear functionals $\Phi^{\times}$ and $\mathcal{H}^{\times}$ respectively.
Since $\mathcal{H}$ is self dual, we obtain%
\begin{equation}
\Phi\subset\mathcal{H}\subset\Phi^{\times}\text{ .}%
\end{equation}

The Nelson operator fully determines the space $\Phi$. However, there are many
inequivalent irreducible representations of an enveloping algebra of a group
characterizing a physical system (e.g. Bohm et al. 1999). Therefore further
restrictions may be required to obtain a realization for $\Phi$, e.g., due to
the convergence properties desired for test functions in $\Phi$. In general
one chooses the weakest topology such that the algebra of operators for the
physical problem is continuous and $\Phi$ is nuclear. The physical symmetries
of the system play an important role in such choices (Bohm et al. 1999).

In RHS QM, the observables form an algebra on the entire space of physical
states (including $\Phi^{\times}$, where Dirac kets reside), so a RHS contains
observables with continuous or even complex eigenvalues, whereas a HS does
not. This means that the basis vector expansion of eigenvectors (Dirac's
spectral decomposition) can be given a rigorous foundation resulting in the
nuclear spectral theorem:
\begin{equation}
|\phi\rangle=\sum_{n}|E_{n})(E_{n}|\varphi)+\int|E\rangle\langle
E|\varphi\rangle d\mu(E)\text{.}%
\end{equation}
Here the rounded bras and kets denote elements elements of $\mathcal{H}$ and
the summation in (5) represents the discrete part of the spectrum. The angular
bras, $\langle\varphi|$, denote elements defined in $\Phi$, while the angular
kets, $|E\rangle$, denote elements defined in $\Phi^{\times}$;\ hence, the
integral in (5) represents the continuous part of the spectrum.

\section{States, Observables and Resonances in Scattering}

A typical scattering experiment consists of an accelerator, which prepares a
projectile in a particular state, a target and detectors. The total
Hamiltonian modeling the interaction of the particle with the target is,
therefore, $H$ = $H_{o}$ + $V$, where $H_{o}$ represents the free particle
Hamiltonian and $V$ the potential in the interaction region. The vectors
representing growing and decaying states are associated with the resonance
poles of the analytically continued S-matrix (Lax and Phillips 1967).

Following the Bohm group, a time arrow emerges in scattering resonances
through imposing the \textit{preparation/registration} arrow of time (Bohm et
al. 1994; Bishop 2004b). The key intuition behind this arrow is that no
observable properties of a state can be measured unless the state has first
been prepared. Following Ludwig (1983; 1985), an in-state of a particular
quantum system (considered as an ensemble of individual systems such as
elementary particles) is prepared by a preparation apparatus (considered
macrophysical). The detector (considered macrophysical) registers so-called
out-states of post-interaction particles. In-states are taken to be elements
$\phi\in\Phi_{-}$ and observables are taken to be elements $\psi\in\Phi_{+}$.
(Resonance states, such as the Dirac, Lippman, Schwinger kets and Gamow
vectors, are elements of $\Phi_{\pm}^{\times}$). This leads to a distinction
between prepared states, on the one hand, and observables, each described by a
separate RHS (Bohm and Gadella 1989; Bohm et al. 1997):%
\begin{subequations}
\begin{align}
\Phi_{-}  &  \subset\mathcal{H}\subset\Phi_{-}^{\times}\text{ }\\
\Phi_{+}  &  \subset\mathcal{H}\subset\Phi_{+}^{\times}\text{ ,}%
\end{align}

\noindent where $\Phi_{-}$ is the Hardy space of the lower complex energy
half-plane intersected with the Schwartz class functions and $\Phi_{+}$ is the
Hardy space of the upper complex energy half-plane intersected with the
Schwartz class functions. As Bohm and Gadella (1989) demonstrate, some
elements of the generalized eigenstates in $\Phi_{-}^{\times}$ and $\Phi
_{+}^{\times}$ correspond to exponentially growing and decaying states
respectively. The semigroups governing these states are\footnote{If $U(t)$ is
a unitary operator on $\mathcal{H}$ and $\Phi\subset\mathcal{H}\subset
\Phi^{\times}$, then $U^{\dagger}$ can be \textit{extended} to $\Phi^{\times}%
$\ provided that (1) $U$ \ leaves $\Phi$ invariant and (2) $U$ \ is continuous
on $\Phi$ with respect to the topology $\tau_{\Phi}$. The operator $U^{\times
}$\ denotes the \textit{extension} of the HS operator $U^{\dagger}$ to
$\Phi^{\times}$\ and is defined by $\langle U\phi|F\rangle=\langle
\phi|U^{\times}F\rangle$ for all $\phi\in\Phi$ and $F\in\Phi^{\times}$. When
the group operator $U^{\dagger}$ is extended to $\Phi^{\times}$, continuity
requirements force the operators $U^{\times}$ to be semigroups defined only on
the temporal half-domains (Bohm and Gadella 1989).}%
\end{subequations}
\begin{subequations}
\begin{align}
\langle\phi|U^{\times}|Z_{R}^{\ast}\rangle &  =e^{-iE_{R}t}e^{\frac{\Gamma}%
{2}t}\langle\phi|Z_{R}^{\ast}\rangle\text{ }t\leq0\text{, }t:-\infty
\rightarrow0\\
\langle\psi|U^{\times}|Z_{R}\rangle &  =e^{-iE_{R}t}e^{-\frac{\Gamma}{2}%
t}\langle\psi|Z_{R}\rangle\text{ }t\geq0\text{, }t:0\rightarrow\infty\text{,}%
\end{align}
where $E_{R}$ represents the total resonance energy, $\Gamma$ represents the
resonance width, $Z_{R}$ represents the pole at $E_{R}-i\frac{\Gamma}{2}$,
$Z_{R}^{\ast}$ represents the pole at $E_{R}+i\frac{\Gamma}{2}$, $|Z_{R}%
^{\ast}\rangle\in\Phi_{-}^{\times}$ represents a growing Gamow vector and
$|Z_{R}\rangle\in\Phi_{+}^{\times}$ represents a decaying Gamow vector. The
$t<0$ semigroup is identified as future-directed along with $|Z_{R}^{\ast
}\rangle$ as a forming/growing state. The $t>0$ semigroup is identified as
future-directed along with $|Z_{R}\rangle$ as a decaying state.

\section{Time-reversed States and Observables}

Following Wigner (1964), the time-reversal operator, $R(t)$, is the HS
representation of the physical spacetime transformation%
\end{subequations}
\begin{equation}
R:(\vec{x},t)\rightarrow(\vec{x},-t)\text{.}%
\end{equation}

\noindent Therefore, $R$ is an element of a co-representation of the extended
Galilei symmetry group (Cari\~{n}ena and Santander 1981) for a nonrelativistic
spacetime (extended Poincar\'{e} group for a relativistic spacetime). These
representations must be unitary and linear except for $R$, which is antilinear.

Wigner originally derived the properties of $R$ for the spacetime symmetry
group extended by time inversions and studied the parity inversion operator
$\Sigma$ and the total inversion operator $T$ in combination with $R$ (Wigner
1964). The parity inversion operator is unitary so its phase can be chosen
such that $\Sigma^{2}$ $=I$ (the identity operator), while $T$ and $R$ are
both anti-unitary, so that the associative law for group multiplication then
dictates that $R^{2}=\varepsilon_{R}I$ and $T^{2}=\varepsilon_{T}I$, where
$\varepsilon_{R}=\pm1$ and $\varepsilon_{T}=\pm1$. The phase of $T$ can be
chosen so that $T=\Sigma R$ (where the order of application of $\Sigma$ and
$R$ is physically immaterial). The extension of the Galilei spacetime symmetry
group is summarized in Table I.\medskip

$\underset{\text{Table I. Properties of the Galilei spacetime symmetry
group.}}{%
\begin{tabular}
[c]{|c|c|c|c|c|}\hline
$\varepsilon_{R}$ & $\varepsilon_{T}$ & $\Sigma$ & $R$ & $T$\\\hline
$(-1)^{2j}$ & $(-1)^{2j}$ & $1$ & $C$ & $C$\\\hline
$-(-1)^{2j}$ & $(-1)^{2j}$ & $\left(
\begin{array}
[c]{cc}%
1 & 0\\
0 & -1
\end{array}
\right)  $ & $\left(
\begin{array}
[c]{cc}%
0 & C\\
-C & 0
\end{array}
\right)  $ & $\left(
\begin{array}
[c]{cc}%
0 & C\\
C & 0
\end{array}
\right)  $\\\hline
$(-1)^{2j}$ & $-(-1)^{2j}$ & $\left(
\begin{array}
[c]{cc}%
1 & 0\\
0 & -1
\end{array}
\right)  $ & $\left(
\begin{array}
[c]{cc}%
0 & C\\
C & 0
\end{array}
\right)  $ & $\left(
\begin{array}
[c]{cc}%
0 & C\\
-C & 0
\end{array}
\right)  $\\\hline
$-(-1)^{2j}$ & $-(-1)^{2j}$ & $\left(
\begin{array}
[c]{cc}%
1 & 0\\
0 & 1
\end{array}
\right)  $ & $\left(
\begin{array}
[c]{cc}%
0 & C\\
-C & 0
\end{array}
\right)  $ & $\left(
\begin{array}
[c]{cc}%
0 & C\\
-C & 0
\end{array}
\right)  $\\\hline
\end{tabular}
\ \ \ }\medskip$

\noindent The index $j$ refers to the spin of the particle being considered
while $C$ is an operator whose $(2j+1)$-dimensional matrix has the elements
$c_{%
\mu
,\nu}=(-1)^{j+%
\mu
}\delta_{%
\mu
,\nu}$, where $-j\leq%
\mu
$ and $\nu\leq j$. In the first representation, where $\varepsilon
_{R}=\varepsilon_{T}=(-1)^{2j}$, there are no changes to the underlying vector
space. This is the typical case discussed in QM (and relativistic quantum
field theory). The other three representations, however, exhibit a doubling of
the vector spaces (note the block matrices in the last three columns of Table
I). In order to track this space doubling, let the index $r=0,1$ label the
rows and columns of the matrices in Table I.

Although no quantum fields have been constructed for representations two and
three of Table I (indeed they are highly problematic), Bohm and co-workers
have constructed models for the fourth representation by applying $R$ to the
states and observables in (7) (Bohm 1995; Bohm and Wickramasekara 1997).
First, consider the growing Gamow vectors for, $\phi^{r=0,\times}\in\Phi
_{-}^{r=0,\times}$. Applying $R$ yields%
\begin{equation}
R\phi^{r=0,\times}=\psi^{r=1,\times}\in\Phi_{+}^{r=1,\times}\text{.}%
\end{equation}
Similarly for the decaying Gamow vectors, $\psi^{r=0,\times}\in\Phi
_{+}^{r=0,\times}$, applying $R$ yields%
\begin{equation}
R\psi^{r=0,\times}=\phi^{r=1,\times}\in\Phi_{-}^{r=1,\times}\text{.}%
\end{equation}
The transformation properties of $R$ may be summarized as $R:\Phi_{\pm
}^{r=0,\times}\rightarrow\Phi_{\mp}^{r=1,\times}$. The temporal evolution of
these time-reversed vectors is also given by semigroups. Identify $r=0$ with
the scattering experiment as normally carried out in the laboratory and $r=1$
with the extended spacetime transformed situation (''time-reversed
counterparts''). Then $U^{\times}(t)\langle\phi,r=0|Z_{R}^{\ast},r=0\rangle
\in\Phi_{-}^{r=0,\times}$, a growing Gamow vector representing a preparable
state for $t\leq0$, is transformed under $R$ into $U^{\times}(-t)\langle
\psi,r=1|Z_{R},r=1\rangle\in\Phi_{+}^{r=1,\times}$, where%
\begin{equation}
e^{iE_{R}t}e^{-\frac{\Gamma}{2}t}\langle\psi,r=1|Z_{R},r=1\rangle
\end{equation}
is restricted to the time domain $t\geq0$ by continuity requirements. In the
case of $|Z_{R}^{\ast},r=0\rangle$, time counts up from $-\infty$ to $0$; in
contrast, for $|Z_{R},r=1\rangle$, time counts down from $\infty$ to $0$,
meaning that it represents a Gamow vector that increases as $t$ decreases.
Similarly, $U^{\times}(t)\langle\psi,r=0|Z_{R},r=0\rangle\in\Phi
_{+}^{r=0,\times}$, a decaying Gamow vector representing observables for
$t\geq0$, is transformed under $R$ into $U^{\times}(-t)\langle\phi
,r=1|Z_{R}^{\ast},r=1\rangle\in\Phi_{-}^{r=1,\times}$, where%
\begin{equation}
e^{iE_{R}t}e^{\frac{\Gamma}{2}t}\langle\phi,r=1|Z_{R}^{\ast},r=1\rangle
\end{equation}
is restricted to the time domain $t\leq0$ by continuity requirements. In the
case of $|Z_{R},r=0\rangle$, time counts up from $0$ to $\infty$; in contrast,
for $|Z_{R}^{\ast},r=1\rangle$, time counts down from $0$ to $-\infty$,
meaning that it represents a Gamow vector that decays as $-t$ increases.

\section{\medskip Matter Meets Mind}

Comparing eqs. (7) with (11) and (12), we can see that in the $r=0$ regime the
association of prepared states with growing eigenvectors and of detected
observables with decaying eigenvectors is quite natural. On the other hand,
the $r=1$ regime has no natural association with physical phenomena (to apply
the eigenstates in this regime ``straightforwardly'' within our framework
would lead to identifying the growing eigenvectors with ``prepared
observables'' and the decaying eigenvectors with ``detected states,''
counterintuitive to say the least).\footnote{The question of interpreting the
time-reversed states and observables was first suggested to me as an
interesting problem by Arno Bohm.}

Suppose we consider an alternative interpretation of the states and
observables of the $r=1$ regime as an abstract representation of mental rather
than material systems. The semigroups in this regime carry vectors from the
future to the past. This could be taken as an abstract representation of final
causation, appropriate to teleological or goal-directed behavior. For example,
suppose I have a particular vision of the kind of person I want to become, say
a more humble person; or suppose I have a particular goal I want to achieve,
say landing a top-flight permanent academic position. These would be examples
of final causation at work in everyday decisions and actions.\footnote{It
would be interesting, though difficult, to connect the framework proposed here
with analyses of goal-directed behavior in cognitive psychology, and cognitive
science more broadly, as the latter perspectives tend to transmute the
apparent final-causal nature of everyday goal-directed behavior into
mechanisms of efficient causation (e.g. Bishop 2005b). Hence, establishing the
desired connection is not straightforward without at least extending the
current cognitive paradigms.} Drawing on the analogy with final causation as a
backwards-directed influence, an eigenvector growing in the backwards time
direction might represent the formation (``preparation'' or ``excitation'') of
such a goal or vision of the future. This could be taken as representing the
building influence of the goal or vision of the future on the present
decision. Similarly, an eigenvector decaying in the backwards time direction
might represent the decision state (``registration'' or ``de-excitation'')
resulting in concrete action toward the goal. It is plausible that decision
states decay back to some kind of ``ready state'' after action is initiated so
that a new decision state can be created for the next set of goals and
actions. The rate of decay could be slower or faster depending on whether the
intended action required more effort of will to ``stay on track'' as it were
to completion or not. The resonance state might be taken as a representation
of the decision itself.

The $r=1$ regime could, then, serve as an abstract model of goal-directed
decision and action. Moreover, both regimes together would play a role in the
abstract description of mental and material systems and their relations. The
$r=0$ regime would correspond to material systems while the $r=1$ regime would
correspond to mental systems. We would, then, have a unified abstract
description of mental-material systems.

Such an abstract description could be deployed to represent a ``dualistic''
distinction between material and mental domains, emerging from a ``monistic''
domain without such a distinction. It has been proposed that this emergence is
related to some temporal symmetry breaking (Atmanspacher 2003; Primas 2004) in
the spirit of ideas of Pauli and Jung (Pauli and Enz 2001), where physical and
psychical aspects originate in a psychophysically neutral domain. The symmetry
breaking envisaged need not be a unique, one-time event, but is perhaps best
understood as an ongoing process due to a number of contingent conditions
giving rise to the mental-material distinction.\footnote{The role of
contingent conditions in the emergence of properties is discussed in (Bishop
and Atmanspacher 2005).} Furthermore, this symmetry breaking can lead to the
Cartesian distinction of the dualistic approach while still allowing for
correlations or forms of interaction emerging from the neutral domain, perhaps
leading to resolution of a number of problems plaguing the dualistic approach.

To be a bit more precise, suppose the neutral domain is characterized by
states $\omega$ and a unitary symmetry (continuum order, automorphic dynamics,
etc.). The dynamics of this domain would then exhibit the time-reversal
symmetry described in section 1 and might be characterized by a one-parameter
unitary group of bounded operators on $\mathcal{H}$. Some, as yet unspecified,
symmetry breaking leads to the generation of time-asymmetric dynamics
characterized by semigroups governing the two regimes, $r=0,1$. As originally
characterized, the states $\omega$ are neutral with respect to mental-material
aspects, whereas after the symmetry breaking, the states are differentiated
into material ($r=0$) and mental ($r=1$) states and processes. It is at the
level of symmetry breaking that the states and observables discussed above
emerge. The characterization of observables in the unitary domain is left
unaddressed here.\footnote{As a referee helpfully pointed out, one could also
consider a more general kind of interpretation of the $r=1$ regime, namely as
representing observational systems, where mental systems are a very important
special case. Space does not permit consideration of this interesting possibility.}

The fact that the $r=0$ and $r=1$ regimes are related to each other via a
time-reversal operator suggests the possibility that there is some form of
intertwining relation among the states and observables of the two regimes. If
so, then the relationship among the elements of the mental and material
domains would not be so starkly disjoint as in Descart\'{e}s' view, where the
two domains are conceived as distinctly different kinds of substances.
Therefore, the distinction between mental and material domains \textit{need
not} imply Descart\'{e}s' metaphysical distinction nor the kinds of
interaction problems encountered in that view.

The RHS framework for QM allows for the description of both time-symmetric and
time-asymmetric phenomena. In particular, it is well-suited for the
description of resonances and other kinds of unstable states. If the
interpretation sketched here makes the time-reversed states and observables of
the $r=1$ regime plausible, then the RHS framework is also well-suited for
such an abstract representation of mental and material states. The unitary
neutral domain might be related to $\mathcal{H}$ while the $r=0$ (material)
and $r=1$ (mental) regimes are related to $\Phi$ and $\Phi^{\times}$.

Although one might wonder about the propriety of using concrete models to
motivate the framework and then subsequently throwing those models to the side
to apply the framework to more abstract questions, this way of proceeding
represents a well-established use of models in mathematical physics (Redhead
1980). There are also technical questions about the application of Wigner's
ideas to observables as well as to semigroup representations, but these
question are fairly straightforward. What is not so straightforward are
questions such as the emergence of time, or the kinds of contingent conditions
leading to the symmetry breaking generating the two regimes. The abstract RHS
framework proposed here appears to be promising as one avenue for exploring
such topics.

\section{Concluding Summary}

One way the fundamental time-reversal invariance of dynamical systems might be
broken is through the presence of resonances and their interactions giving
rise to unstable dynamical systems. When time-reversal invariance is broken,
this results in two well-defined time arrows associated with semigroups
bearing time orientations, one directed toward the future and one directed
toward the past. Scattering resonances provide an example where time-reversal
invariance is broken. The resulting forward-directed semigroups and states
correspond to the processes of resonance formation, decay and detection, but
the backward-directed semigroups and states are thought to have perhaps only
mathematical significance. Here, I have sketched a possible interpretation of
these latter semigroups and states as corresponding abstractly to the domain
of mental systems, while the forward-directed semigroups and states would
correspond to the domain of material systems. The crucial idea is that these
two domains might emerge from a more fundamental domain that is neutral with
respect to any mental-material distinction and that, hence, various
possibilities exist for relations between the emergent domains that are
typically precluded by traditional Cartesian dualisms. The RHS framework seems
well-suited for describing and exploring these possibilities.

\begin{acknowledgement}
I would like to thank H. Atmanspacher and A. Bohm for helpful discussions.
Financial support from the Alexander von Humboldt Foundation as well as the
Federal Ministry of Education and Research and the German government's Program
for the Investment in the Future of the are gratefully acknowledged.
\end{acknowledgement}

\section{References}

\noindent I. Antoniou and S. Tasaki, \textit{Int. J. Quant. Chem.},
\textbf{46} (1993) 425

\noindent I. Antoniou and I. Prigogine, \textit{Physica A}, \textbf{192
}(1993) 443.

\noindent H. Atmanspacher, \textit{BioSystems}, \textbf{68} (2003) 19.

\noindent R. C. Bishop, \textit{Studies in History and Philosophy of Modern
Physics} \textbf{35} (2004a) 1.

\noindent R. C. Bishop, \textit{Int. J. Theor. Phys.}, \textbf{43} (2004b) 1675.

\noindent R. C. Bishop, \textit{Int. J. Theor. Phys.}, (2005a) in press.

\noindent R. C. Bishop, ``Cognitive Psychology: Hidden Assumptions,'' in
\textit{Critical Thinking About Psychology: Hidden Assumptions and Plausible
Alternatives}, B. Slife, J. Reber and F. Richardson (eds.), Washington, D. C.:
American Psychological Association Books, (2005b) 151.

\noindent R. C. Bishop and H. Atmanspacher, (2005), ``Contextual Emergence in
the Description of Properties,'' submitted.

\noindent A. Bohm, \textit{Phys. Rev. A}, \textbf{51} (1995) 1758.

\noindent A. Bohm, I. Antoniou and P. Kielanowski, \textit{Phys. Lett. A},
\textbf{189} (1994) 442.

\noindent A. Bohm and M. Gadella, \textit{Dirac Kets, Gamow Vectors, and
Gel'fand Triplets, Lecture Notes in Physics, vol. 348}, Springer, Berlin, (1989).

\noindent A. Bohm, M. Gadella and S. Wickramasekara, ``Some Little Things
about Rigged Hilbert Spaces and Quantum Mechanics and All That,'' in I.
Antoniou and G. Lumer (eds.) \textit{Generalized Functions, Operator Theory,
and Dynamical Systems}. Boca Raton, FL: Chapman \& Hall/CRC, (1999) 202.

\noindent A. Bohm, S. Maxson, M. Loewe and M. Gadella, \textit{Physica A},
\textbf{236} (1997) 485.

\noindent A. Bohm and S. Wickramasekara, \textit{Found. of Phys.}, \textbf{27}
(1997) 969.

\noindent J. Cari\~{n}ena and M. Santander, \textit{J. Math. Phys.},
\textbf{22} (1981) 1548.

\noindent P. Lax and R. Phillips, \textit{Scattering Theory},\textit{
}Academic Press,\textit{ }New York, (1967).

\noindent G. Ludwig, \textit{Foundations of Quantum Mechanics, Vol. I},
Springer, Berlin, (1983).

\noindent G. Ludwig, \textit{Foundations of Quantum Mechanics, Vol. II},
Springer, Berlin, (1985).

\noindent E. Nelson, \textit{Ann. Math.}, \textbf{70} (1959) 572.

\noindent W. Pauli and C. P. Enz, \textit{Atom and Archetype: The Pauli/Jung
Letters, 1932-1958}. Princeton: Princeton University Press, (2001).

\noindent H. Primas, \textit{Mind and Matter}, \textbf{1}, (2004) 81.

\noindent M. Redhead, \textit{Brit. J. Phil. Sci.}, \textbf{31}, (1980) 145.

\noindent F. Treves, \textit{Topological Vector Spaces, Distributions and
Kernels}. New York: Academic Press, (1967).

\noindent E. Wigner, ``Unitary Representations of the Inhomogeneous Lorentz
Group Including Reflections,'' in \textit{Group Theoretical Concepts and
Methods in Elementary Particle Physics}, F. G\"{u}rsey (ed.), Gordon and
Breach, Science Publishers, New York, (1964).
\end{document}